\documentclass[%
 reprint,
 amsmath,amssymb,
 aps,
prb,
]{revtex4-2}

\usepackage{graphicx}
\usepackage{dcolumn}
\usepackage{bm}
\usepackage{multirow}
\usepackage{booktabs}
\usepackage{makecell}
\usepackage{graphicx}

\begin{document}

\title{Nonsymmorphic symmetry adapted finite element modeling of glide-symmetric photonic structures}

\author{Lida Liu}
\affiliation{School of Optical and Electronic Information, Huazhong University of Science and Technology, Wuhan}

\author{Jingwei Wang}
\affiliation{School of Optical and Electronic Information, Huazhong University of Science and Technology, Wuhan}

\author{Yuhao Jing}
\affiliation{School of Optical and Electronic Information, Huazhong University of Science and Technology, Wuhan}

\author{Songzi Lin}
\affiliation{School of Optical and Electronic Information, Huazhong University of Science and Technology, Wuhan}

\author{Zhongfei Xiong}
\email{xiongzf94@outlook.com}
\affiliation{School of Optical and Electronic Information, Huazhong University of Science and Technology, Wuhan}

\author{Yuntian Chen}
\email{yuntian@hust.edu.cn}
\affiliation{School of Optical and Electronic Information, Huazhong University of Science and Technology, Wuhan \\ Optics Valley Laboratory, Wuhan \\ Wuhan National Laboratory of Optoelectronics, Huazhong University of Science and Technology, Wuhan}%

\date{\today}

\begin{abstract}
Space group theory is pivotal in the design of nanophotonics devices, enabling the characterization of periodic optical structures such as photonic crystals. The aim of this study is to extend the application of nonsymmorphic space groups in the field of numerical analysis for research and design of nanophotonics devices. In this work, we introduce the nonsymmorphic symmetry adapted finite element method, and provide a systematic approach for efficient band structure analysis of photonic structures with nonsymmorphic groups. We offer a formal and rigorous treatment by specifically deriving the boundary constraint conditions associated with the symmetry operations and their irreducible representations and decomposing the original problem into different subtasks. our method fully accounting for non-primitive translations and nonstructural symmetries like time-reversal symmetry and hidden symmetries. We demonstrate the effectiveness of our method via computing the band structure of photonic structures with a layer group, a plane group, and a space group. The results exhibit excellent agreement with those obtained using the standard finite element method, showcasing improved computational efficiency. Furthermore, the decomposition of the original problem facilitates band structure classification and analysis, enabling the identification of the different bands among the band structure in various subtasks. This advancement paves the way for innovative designs in nanophotonics.
\end{abstract}

\maketitle


\section{\label{sec:level1}INTRODUCTION\protect\\}

Space group theory plays a significant role in photonics\cite{WOS:000453254900011,WOS:001293426500003,WOS:000454637100015}, condensed matter physics\cite{WOS:000404332000042,WOS:000443154900001,WOS:000618283700013}, and numerical computation\cite{photonics10060691,Wang2024GeneralizedBB,10776583,Wang:24} by fundamentally describing how symmetry affects the interactions of light fields with matter. In recent years, with the development of topological optics, there has been increasing interest in how to use symmetry to construct periodic optical structures such as photonic crystals and metasurfaces with novel characteristics, where space groups effectively characterize their properties. In group theory, space groups are categorized into two types: symmorphic and nonsymmorphic, each with distinct properties. Symmorphic groups can induce topologically protected photonic states\cite{WOS:000943801100001,WOS:000385588900003,2016Symmetry} and enhance optical nonlinearities\cite{WOS:000453254900011,WOS:000996342500001} as well as other topological phenomena. Nonsymmorphic groups, in particular, are more complex as they incorporate operations that combine point groups with non-primitive translation, such as glide reflections and screw rotations\cite{https://doi.org/10.1107/S0567739473001476}. Nonsymmorphic groups have opened up a rich field for intriguing topological phenomena. For example, when combined with time-reversal symmetry, they can lead to Kramer-like optical degeneracy. Meanwhile, they play an important role in the discovery of new phases, such as topological insulators\cite{2020Antiferromagnetic,2020Superconductivity,2024Hidden,2022Nonsymmorphic}, and topological semimetals \cite{WOS:001255332900003,WOS:000752108100006,WOS:000886677900001}. Furthermore, they can facilitate and protect the emergence of exotic fermionic quasiparticles, including Dirac, Möbius, and hourglass fermions\cite{doi:10.1142/9789811264153_0034}. Moreover, the introduction of symmetry protection through nonsymmorphic symmetries results in distinct characteristics such as band sticking and crossings\cite{PhysRevB.105.064517, WOS:000770623900004}. The differences between nonsymmorphic groups and symmorphic groups have greatly enriched the design of photonic devices, such as photonic crystals\cite{WOS:001065505200001,WOS:001209583100001,Adam2008Spectral}, metamaterials\cite{WOS:001000848200001, WOS:000790993900001}, waveguides\cite{WOS:000796530400018,WOS:001215536202155,WOS:000795897700002}, and topological couplers \cite{Bazhan2021JosephsonOO}, which leverage these symmetries to achieve extraordinary performance.

\begin{figure*}
\includegraphics{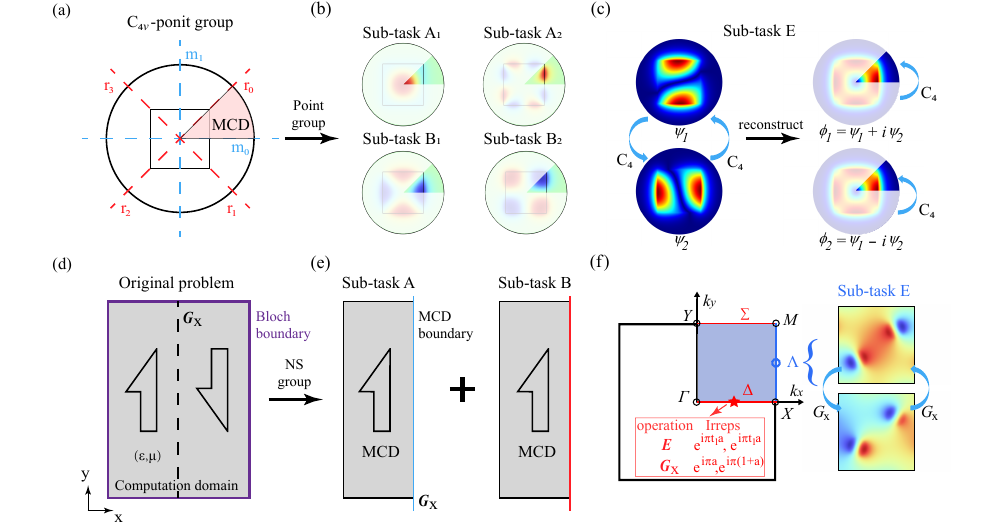}
\caption{(a) Modal analysis on the $C_{4v}$ structure. The original problem are decomposed into five sub-tasks. (b) The non-degenerate modes computed by sub-tasks $A_1,B_1,A_2,B_2$; (c) The symmetry-protected degenerate modes computed by sub-task $E$. (d) Standard FEM for band analysis on the $p1g1$ structure, gliding reflection axis denoted as $G_{x}$; (e) Decomposed sub-tasks with its own MCD; (f) FBZ and irreducible FBZ (blue region) in $\boldsymbol{k}$-space along with corresponding high symmetry points/lines. The nonsymmorphic symmetry-protected degenerate modes computed by sub-task $E$.}
\label{fig0}
\end{figure*}

Space groups are invaluable not only for theoretical analysis of the unique properties of optical devices but also for simplifying their computational analysis. For instance, the minimum computation domain (MCD) is usually restricted to a single unit cell in real space by using the Bloch boundary condition, and band structure analysis can also only be considered in the irreducible first Brillouin zone (FBZ). Furthermore, systematic approaches have been developed to leverage point groups to reduce computational domains for structures exhibiting symmorphic symmetries\cite{photonics10060691,Wang2024GeneralizedBB,10776583}. As shown in Fig. \ref{fig0}(a-c). Using the point group, only one-eighth of the entire domain is modeled as the MCD (highlighted by the magenta region), and decomposes the original problem into five sub-tasks\cite{Wang:24}. Moreover, recent studies have applied glide and screw symmetries to decrease computational costs by focusing on the minimal subunit cell rather than the entire cell. For example, a glide-symmetric structure can be equivalently transformed into a non-glide-symmetric structure with a reduced period\cite{8728174}. Furthermore, research has employed multiple waveguide modes to derive the dispersion diagram of a line using a multi-modal transmission matrix\cite{sym11050620}. However, a systematic approach that reduces the computational domains of periodic structures using nonsymmorphic symmetry  adapted finite element method still remain elusive.

In previous work, computations for degenerate modes were achieved using point groups, as the influence of lattice was not considered\cite{photonics10060691,Wang2024GeneralizedBB,10776583,Wang:24}. However, as the non-primitive translations in the Bravais lattice is included, different momentum positions in reciprocal space exhibit distinct irreducible representations, as shown in Fig. \ref{fig0}(f)\cite{ZHANG2019244}. In point group, the degenerate modes $\psi_{1,2}$ computed by sub-task $E$ corresponding to the second-order representation, can be fully decoupled by reconstructing another pair of degenerate modes $\phi_{1,2}$ into two sub-task via basis rotation under $C_4$ symmetry as depicted in Fig. \ref{fig0}(c).  In nonsymmorphic groups such as the $p1g1$ group shown in Fig. \ref{fig0}(d), the rectangular unit cell contains a gliding reflection operation denoted as $G_{x}$ along the dashed vertical line. As shown in Fig. \ref{fig0}(e), to calculate the band structure of $\Gamma$-$X$ in $\boldsymbol{k}$-space, our NSA-FEM decomposes the original problem into two sub-tasks, only models half of the entire domain, each satisfies different MCD boundary. Because operations $E$, i.e., integer lattice vector translation and $G_x$ both have second-order representation at $\Lambda$, it is not possible to reconstruct another pair of decoupled degenerate modes, as indicated by the entanglement of the two degenerate modes shown in Fig. \ref{fig0}(f). Moreover, band degeneracy in non-symmorphic space groups may not always be fully described by space groups\cite{2020Hidden}, on one hand, time-reversal symmetry contributes to band degeneracies at band edges; on the other hand, there may also be hidden symmetries. Therefore, using a point group alone cannot accomplish the calculation of degenerate modes in a nonsymmorphic group. Additionally, various Bravais lattices form groups of differing dimensions, including plane groups, space groups, and sub-periodic groups. Developing a unified approach to handle nonsymmorphic groups across these different dimensions remains a significant challenge. Indeed, there is a notable gap in the application of space group theory to effectively utilize nonsymmorphic symmetries in reducing the complexity of problems such as searching for new states of matter, discovering novel topological phenomena, and computing band structures.

In this paper, we provide a systematic study on the symmetry adapted finite element method applied to photonic structures with the nonsymmorphic space symmetry, and employ this method to calculate the band structure and the associated eigen modal field. This method takes advantage of both symmorphic and nonsymmorphic groups to simplify computational complexity. The eigenmodes are classified by using the character tables from group theory, which incorporates the non-primitive translations associated with \(\boldsymbol{k}\)-vectors at different locations. Moreover, taking into account the effect of nonstructural symmetries, i.e., time reversal symmetry,  we consider irreducible representations of the full-group induced by $\boldsymbol{k*}$. Additionally, space groups with different dimensions can be equally described by character tables. As such, the original modal problem is decomposed into several decoupled sub-tasks, each of which corresponds to different symmetry types, as shown in Fig. \ref{fig0}(d-f). Once these sub-tasks are solved, the original problem can be fully recovered. Compared to the standard finite element method (FEM), each completely decoupled sub-task in our method involves fewer degrees of freedom (DOFs) and also facilitates the implementation of parallel computation within certain algorithmic framework. Thereby, the total computational time could be reduced significantly. In summary, our work fundamentally extends the symmetry adapted FEM framework based on space group theory, with the capability of speeding up the computation of any complex 3D photonic crystals exhibiting nonsymmorphic group symmetries.

\section{THEORY AND METHODS}
\subsection{General framework}

For ease of description, we consider the structure shown in Fig. \ref{fig0}(d), which illustrates the lattice structure of a photonic crystal. This structure belongs to the rectangular lattice category. The plane group index of the photonic crystal grating is p1g1 (number 4), possessing only a single glide operation, namely $G_{x}=\left\{ M_{y} \middle| \boldsymbol{\tau} \right\}$. We use the Seitz operator\cite{1996Group} denote a space group operation, where $G_{x}$ represents a glide operation with the non-primitive translation along the x-axis, $M_{y}$ represents a mirror operation with the mirror plane oriented along the y-axis, and $\boldsymbol{\tau} = \frac{\alpha_{1}}{2}\hat{x}$ represents the non-primitive glide translation vector. 

In the standard FEM, the computational domain encompasses the entire periodic unit cell, and the electric field satisfies the source-free vector wave equation\cite{1998Quick}
\begin{equation}\label{eqn1} 
  \nabla\times\overline{\mu_r}^{-1}\nabla\times \boldsymbol{E}(\boldsymbol{r})-k_0^2\overline{\epsilon}_r\boldsymbol{E}(\boldsymbol{r})=0,
\end{equation}
where $\nabla \times$ denotes the curl operator, $\overline{\mu_r}$/$\overline{\epsilon_r}$ is the relative permeability(permittivity) tensor, and $k_0$ is the vacuum wave number. The unit cell boundary (indicated by the purple solid line in Fig. \ref{fig0}(d)) employs the Bloch boundary conditions to effectively simulate the periodic structure, which satisfy the equation
\begin{equation}\label{eqn2} 
 \boldsymbol{E}^{k}(\boldsymbol{r}) = e^{i\boldsymbol{k} \cdot \boldsymbol{R}_{n}}\boldsymbol{E}^{k}\left( {\boldsymbol{r} + \boldsymbol{R}_{n}} \right),
\end{equation}
where $\boldsymbol{R}_{n}$ represents the primitive translation transformations. Moreover, the concept of the Brillouin zone is as important in the calculation of the band structure of photonic crystals as the phase factor $e^{i\boldsymbol{k} \cdot \boldsymbol{R}_{n}}$ of the Bloch boundary conditions are related to $\boldsymbol{k}$ which represents the wave vector. A Brillouin zone is defined as a Wigner-Seitz primitive cell in the reciprocal space ($\boldsymbol{k}$-space). FBZ is the smallest volume entirely enclosed by planes that are the perpendicular bisectors of the reciprocal lattice vectors drawn from the origin\cite{1968Theory}. The smallest region within the Brillouin zone where $\boldsymbol{k}$ are not related to symmetry is called the irreducible FBZ. Fig. \ref{fig0}(f) shows the FBZ and the high-symmetry lines/points in the $\boldsymbol{k}$-space, the blue region shows irreducible FBZ. Typically, band structures and modes are analyzed only within the irreducible FBZ.

Considering the influence of the Bravais lattice, nonsymmorphic symmetries exhibit distinct irreducible representations (Irreps) at different $\boldsymbol{k}$-vectors because of the non-primitive translation. For example, along the high symmetry momentum lines, such as at location $\Delta$, which Irreps can be obtained by using the normal-subgroup induction method described in Appendix A\cite{10.1093/oso/9780199582587.001.0001,ZHANG2019244}. The Irreps corresponding to the operations are depicted in Fig. \ref{fig0}(f). The operation $E={\{1|t_1,t_2\}}$ corresponds to the Bloch boundary condition. By employing the nonsymmorphic symmetry operation $G_x=\{m_{01}|\frac{1}{2},0\}$, we can get the other half of the structure from one half of the structure by the transformation $G_x$. Thus, we can modal only half of the structure, referred to as the MCD, as illustrated in Fig. \ref{fig0}(e). The original problem corresponding to standard FEM is decomposed into decoupled(multiple) sub-tasks, each associated with a class of the symmetry group. In addition to the boundary constraints on the original outer boundary, each sub-task must also satisfy the boundary constraints related to the symmetry constraints on the MCD boundary. We refer to this method as nonsymmorphic symmetry adapted FEM (NSA-FEM).

\begin{figure}
\includegraphics{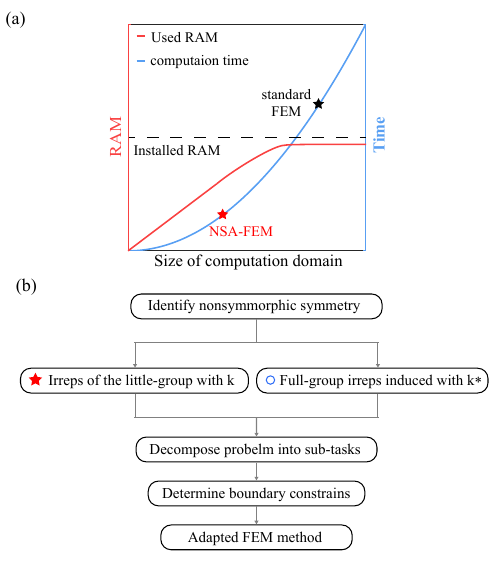}
\caption{\label{fig:epsart} (d) As the size of computational domain varies, the RAM and computational time of the FEM also fluctuate. (e) The general procedure of the NSA-FEM.}
\label{fig1}
\end{figure}

Although the NSA-FEM involves multiple sub-tasks, the size of the computational domain are significantly reduced. As shown in Fig. \ref{fig1}(a), due to the nonlinear relationship between computation time and the size of the computational domain, this method significantly improves computational efficiency for larger-scale problems by reducing the size of the computational domain and the number of modes associated with each decoupled sub-task. Moreover, this improvement becomes particularly significant as the random-access memory(RAM) approaches the computer's installed RAM. Furthermore, the complete decoupling of these sub-tasks allows for enhanced computational efficiency through parallel computing techniques. Consequently, the NSA-FEM markedly improves computational efficiency compared to standard FEM.

The implementation procedure of NSA-FEM is shown in Fig. \ref{fig1}(b), divided into five steps.
 \begin{enumerate}
    \item Identify the nonsymmorphic symmetry of the calculated structure and determine the irreducible FBZ.
    \item Select the Irreps of the little group at \(\boldsymbol{k}\) or the full-group Irreps induced at \(\boldsymbol{k}*\) based on the symmetry properties and degeneracy.
    \item Decompose the original problem into sub-tasks based on the selected character table.
    \item Determine the symmetry constraints for each sub-task from the selected symmetry properties, and derive the boundary constraints that can be applied to FEM.
    \item Apply the corresponding boundary constraints to the sub-tasks and solve them to obtain the band of different symmetry types.
\end{enumerate}

For Steps 1 and 2, we can first determine the generators based on the structure. The generators are a minimal set of symmetry operations that can be used to generate all other symmetry operations. Through the generators, we can look up to get the corresponding nonsymmorphic group, and further look up to determine its irreducible FBZ. Once the nonsymmorphic group and irreducible FBZ of the computational structure are determined, the Irreps associated with them can be identified. Through this group, it is possible to obtain the Irreps of the little group, as well as the symmetry operations at different high-symmetry momentum points (k-vectors) that lie on the boundaries of the irreducible FBZ. These symmetry operations are utilized to truncate the computational domain and decompose the original problem into sub-tasks according to the Irreps. In special cases, such as mode degeneracy due to time-reversal symmetry, the symmetry constraints can also be obtained using the full-group Irreps induced from \(\boldsymbol{k}*\). The details will be discussed in Section 2.2. For Steps 3 and 4, the different classes within the Irreps have different characters, leading to variations in the MCD boundary conditions. This results in the decomposition of the original problem into several sub-tasks associated with the classes of the Irreps, which will be discussed in Section 2.3. For Step 5, the application of FEM to handle the matrix of eigenvalue problem under nonsymmorphic symmetry constraints, which are based on the derived boundary conditions, will be discussed in Section 2.4.

\subsection{Irreps induced with \(\boldsymbol{k}\) or \(\boldsymbol{k}*\)}
The 2D plane group can be seen as a projection of the 3D space group. All generators and general positions of the space groups as well as the Irreps of the space groups can be referred to in the Bilbao Crystallographic server\cite{2020SpaceGroupIrep}. For instance, the p1g1 plane group can be equivalently regarded as a particular surface within the space group Pc (number 7) and can be referred to in the Bilbao Crystallographic server. The Irreps of each high symmetry \(\boldsymbol{k}\)-vector and the corresponding symmetry operations are tabulated in Table \ref{tab:table1}, where the location is denoted by reciprocal lattice vectors \(\boldsymbol{b_1}\) and \(\boldsymbol{b_2}\)  as \(\boldsymbol{k}=(m,n)=m\boldsymbol{b_1}+n\boldsymbol{b_2}\).

\begin{table}
\caption{\label{tab:table1}Irreps of the high symmetry \(\boldsymbol{k}\)-vector for plane group p1g1}
\begin{ruledtabular}
\renewcommand\arraystretch{1.4}
\begin{tabular}{c|ccc|c}
Location&\makecell{Seitz \\ Symbol}&\multicolumn{2}{c|}{Representation} &\makecell{Irreps \\ origin}\\ \hline
\multirow{2}{*}{$\Gamma$:(0,0)}&\(\{1|t_1,t_2\}\)&1&1&\multirow{6}{*}{$\boldsymbol{k}$} \\ 
&\(\begin{array} {c}\{\mathfrak{m}_{01}|\frac{1}{2},0\} \end{array}\)&1&-1&\\  \Xcline{1-4}{0.05pt}
\multirow{2}{*}{\(\Delta\):($a$,0)\footnote{$0<a <\frac{1}{2}$}}&\(\{1|t_1,t_2\}\)& \(e^{i2\pi t_1a}\) & \(e^{i2\pi t_1a}\)& \\
&\(\begin{array} {c}\{\mathfrak{m}_{01}|\frac{1}{2},0\} \end{array}\)&\(e^{i\pi a}\)&\(e^{i\pi (1+a)}\)&\\ \Xcline{1-4}{0.05pt}
\multirow{2}{*}{\(X\):(\(\frac{1}{2}\),0)}&\(\{1|t_1,t_2\}\)& \(e^{i\pi t_1}\) & \(e^{i\pi t_1}\)& \\
&\(\begin{array} {c}\{\mathfrak{m}_{01}|\frac{1}{2},0\} \end{array}\)&\(i\)&\(-i\)&\\ \Xcline{1-5}{0.05pt}
\multirow{2}{*}{\(\Lambda\):(\(\frac{1}{2}\),\(a\))}&\(\{1|t_1,t_2\}\)&\multicolumn{2}{c|}{\(\left.\left(
\begin{array}
{cc}e^{i\pi(t_1+2t_2a)} & 0 \\
0 & e^{i\pi(t_1+2t_2a)}
\end{array}\right.\right)\)}&\multirow{2}{*}{$\boldsymbol{k*}$}  \\
&\(\begin{array} {c}\{\mathfrak{m}_{01}|\frac{1}{2},0\} \end{array}\)&\multicolumn{2}{c|}{\(\left.\left(
\begin{array}
{cc}0 & 1 \\
-1 & 0
\end{array}\right.\right)\)}& \\ \Xcline{1-5}{0.05pt}
\multirow{2}{*}{\(M\):(\(\frac{1}{2}\),\(\frac{1}{2}\))}&\(\{1|t_1,t_2\}\)&\(e^{i\pi(t_{1+}t_2)}\)&\(e^{i\pi(t_{1+}t_2)}\)&\multirow{6}{*}{$\boldsymbol{k}$} \\
&\(\begin{array} {c}\{\mathfrak{m}_{01}|\frac{1}{2},0\} \end{array}\)&\(i\)&\(-i\)&\\ \Xcline{1-4}{0.05pt}
\multirow{2}{*}{\(\Sigma\):(\(a\),\(\frac{1}{2}\))}&\(\{1|t_1,t_2\}\)& \(e^{i\pi(2t_1a+t_2)}\) & \(e^{i\pi(2t_1a+t_2)}\)& \\
&\(\begin{array} {c}\{\mathfrak{m}_{01}|\frac{1}{2},0\} \end{array}\)&\(e^{i\pi a}\)&\(e^{i\pi (1+a)}\)&\\ \Xcline{1-4}{0.05pt}
\multirow{2}{*}{\(Y\):(0,\(\frac{1}{2}\))}&\(\{1|t_1,t_2\}\)& \(e^{i\pi t_2}\) & \(e^{i\pi t_2}\)& \\
&\(\begin{array} {c}\{\mathfrak{m}_{01}|\frac{1}{2},0\} \end{array}\)&\(1\)&\(-1\)&\\
\end{tabular}
\end{ruledtabular}
\end{table}

From the fact that any element of the group \(\{{R}|\boldsymbol{\tau}+\boldsymbol{R}_n\}\) can be expressed as a combination of a translation \(\{{E}|\boldsymbol{R}_n\}\) and a coset representative \(\{{R}|\boldsymbol{\tau}\}\) as presented in Eq.~(\ref{apeqn1}), it follows that Irrep matrices of \(\{{R}|\boldsymbol{\tau}+\boldsymbol{R}_n\}\) are equal to products of the type \(D^\mathrm{k}(\{\mathrm{R}|\boldsymbol{\tau}+\boldsymbol{R}_n\})=D^\mathrm{k}(\{\mathrm{E}|\boldsymbol{R}_n\})D^\mathrm{k}(\{\mathrm{R}|\boldsymbol{\tau}\})\). Thus, in Table \ref{tab:table1}, the first row of every \(\boldsymbol{k}\)-vector's Seitz symbol corresponds to a general translation \(\{{E}|\boldsymbol{R}_n\}\) while the second row shows the data of the representative \(\{{R}|\boldsymbol{\tau}\}\) of the coset decomposition of the little group with respect to its translation subgroup. Furthermore, the sub-table composed of the operations under the Seitz symbol and their representations corresponds to the character table\cite{Xiong:20}. The operations form the classes, and the representations are the allowed representations obtained from the isomorphic group. For more details, see Appendix A.

In Fig. \ref{fig0}(f), the red boundary indicates that the \(\boldsymbol{k}\)-vector at this boundary exhibits glide symmetry, whereas the blue boundary signifies the absence of such symmetry. 

On the red boundary, the Irreps are induced from the allowed Irreps of the little group $^dG^{\boldsymbol{k}}$. The normal-subgroup induction method for the calculation of space groups Irreps was described in Appendix A. The main steps of the induction involve the identification of the second kind subgroups $G_k$ and the quotient group $G_k/T_k$, the determination of the corresponding little groups and the allowed representations, and finally, induce the Irreps of the space group from the allowed representations.

In the aforementioned discussion of the red boundary, the time-reversal symmetry is not included. In contrast, on the blue boundary, the time-reversal symmetry exists and plays a significant role. Thus, we need to consider the Irreps with respect to complex conjugation, or as it is commonly referred to, with respect to 'reality'. Once the reality of Irrep has been determined and, for the complex Irreps, the pairs of conjugated Irreps have been identified, the time-reversal (TR)-invariant irreps can be constructed. Furthermore, the Irreps are induced with the arms of the star $\boldsymbol{k}*$, which precedes the table of the matrices of the full-group Irreps $D^{\boldsymbol{k}*,j}$ of $^{d}G$ induced from the allowed irreps $D^{\boldsymbol{k},j}$ of the little group $^{d}G^k$. In summary, we can employ the full-group Irreps to jointly solve for the arms of the star ( \(k_1^*=(m,n)\) ) and ( \(k_2^*=(m,-n)\) ), thereby obtaining a set of degenerate solutions; see details in Appendix B\cite{Elcoro:ks5574}.

\subsection{Problem decomposition and the symmetry constraint of boundary condition for each sub-task}
\begin{figure*}
\includegraphics{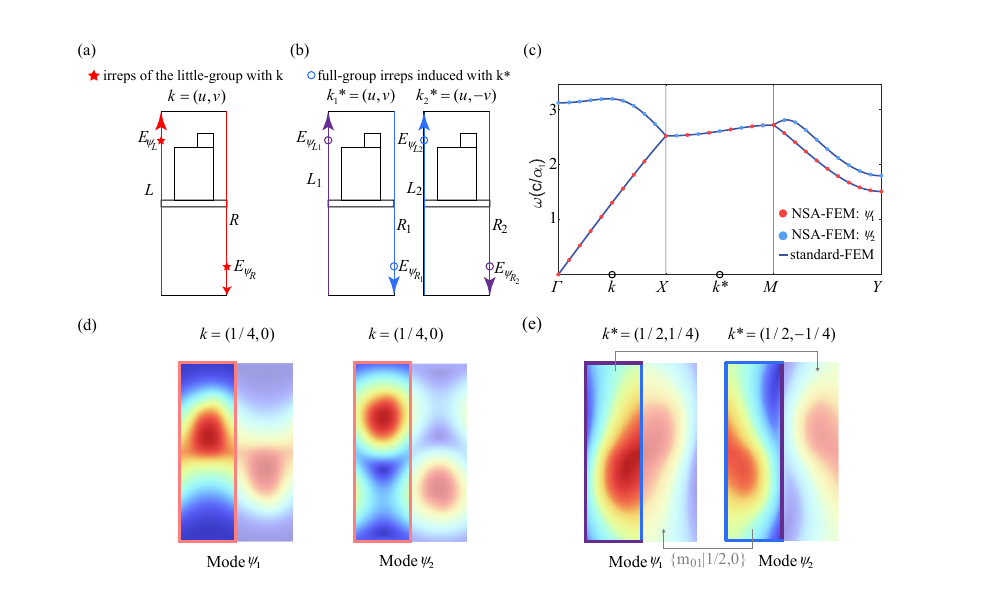}
\caption{(a) Symmetry constraints corresponding to the Irreps of the little group associated with \(k\). The original rectangular unit cell comprises two gratings and a substrate. (b) Symmetry constraints corresponding to the full-group Irreps induced with k*. (c) Band structure of the photonic crystal grating in the rectangular lattice. (d) Non-degenerate modes \(\psi_i\) (e) degenerate modes \(\psi_{1,2,...,i}\) constrained by operation $\{m_{01}|\frac{1}{2},0\}$.}
\label{fig2}
\end{figure*}
Once the Irreps of the symmetry group is determined, the corresponding character table can be used to decompose the original problem. For instance, as shown in Table \ref{tab:table1}, the glide symmetry categorizes all modes into two different symmetry types. Using the high symmetry momentum \(\Delta\) and \(\Lambda\) as an example, one is able to  decompose the original problem  into two decoupled sub-tasks, since the two types of representations at \(\Delta\)  are one-dimensional (1D). In contrast, the representation at \(\Lambda\) is two-dimensional (2D), which involves two sub-tasks and the degeneracy of two different symmetry types, each corresponding to a distinct symmetry modal type. This decomposition inherently facilitates band structure classification and analysis, enabling the identification of the different bands in various sub-tasks prior to the completion of computations. By definition, applying the Seitz operator to a vector field \(E(\boldsymbol{r})\) will result in the following transformation,
 \begin{equation}\label{eqn4} 
P(\{R|\boldsymbol{\tau+R_n}\}))\boldsymbol{E_{\psi}}\boldsymbol{(r)}=D^{kj}(\{R|\boldsymbol{\tau}+\boldsymbol{R}_n\})\boldsymbol{E_{\psi}}\boldsymbol{(r)},
\end{equation}
\begin{equation}\label{eqn5} 
P(\{R|\boldsymbol{\tau}+\boldsymbol{R}_{n}\})\boldsymbol{E}_{\boldsymbol{\psi}}(\boldsymbol{r})=\{R|\boldsymbol{\tau}+\boldsymbol{R}_{n}\}\boldsymbol{E}_{\boldsymbol{\psi}}(\{R|\boldsymbol{\tau}+\boldsymbol{R}_{n}\}^{-1}\boldsymbol{r}),
\end{equation}
where \(\boldsymbol{E_{\psi}(r)}\) is the electric field of mode \(\psi\). \(D^{kj}(\{R|\boldsymbol{\tau}+\boldsymbol{R}_n\})\) is the matrix form of irreducible representation corresponding to the symmetry operation \((\{R|\boldsymbol{\tau}+\boldsymbol{R}_n\})\), which describes the constraints imposed by symmetry on modes of different sub-tasks. The trace of \(D^{kj}(\{R|\boldsymbol{\tau}+\boldsymbol{R}_n\})\), known as the character, is denoted as \(\chi(\{R|\boldsymbol{\tau+R_n} \})\). This character can be found in Table \ref{tab:table1}, with data sourced from the Bilbao Crystallographic server. Using Eq.~(\ref{eqn4}), we can derive the boundary constraints of MCD based on the constraints of symmetry \((\{R|\boldsymbol{\tau}+\boldsymbol{R}_n\})\). There are two situations outlined as follows.
\begin{itemize}
    \item Method 1:  If symmetry operation  $\{R|\boldsymbol{\tau+R_n}\}$  acts on non-degenerate modes \(\psi_i\), \(\chi(\{R|\boldsymbol{\tau+R_n}\})\) is reduced  into  a scalar, and \(\boldsymbol{E}_{\boldsymbol{\psi}}(\boldsymbol{r})={E}_{\psi_i}(\boldsymbol{r})\);
    \item Method 2:  If  symmetry operation $\{R|\boldsymbol{\tau+R_n}\}$  acts on degenerate modes \({\psi_{1,2,...,n}}\), \(\chi(\{R|\boldsymbol{\tau+R_n}\})\) is a matrix and \(\boldsymbol{E}_{\boldsymbol{\psi}}(\boldsymbol{r})=[{E}_{\psi_1}(\boldsymbol{r}),{E}_{\psi_2}(\boldsymbol{r}),...,{E}_{\psi_n}(\boldsymbol{r})]\).
\end{itemize}
Consequently, the modal symmetry properties of high symmetry momenta \(\Delta\) and \(\Lambda\) can be categorized into Method 1 and Method 2, respectively, i.e.,  \( \Delta \) for Method 1 and \(\Lambda\) for Method 2. We employ the glide symmetry \({\{m_{01}|1/2,0\}}\) alongside Eq.~(\ref{eqn2}), and can derive the symmetry constraints on the MCD boundary at $\Delta$ as follows, 
 \begin{equation}\label{eqn6} 
E_\Delta(r_R)=\pm e^{i\pi a}E_\Delta(r_L). 
\end{equation}
Concretely, the MCD boundary L and R shown in Fig. \ref{fig2}(a) are denoted  by \(\boldsymbol{r}_L\) and \(\boldsymbol{r}_R\). And the symmetry constraint on the MCD boundary at $\Lambda$ are given as follows,
 \begin{subequations}\label{eqn7} 
 \begin{align}
  E_{\Lambda_{2}}(\boldsymbol{r}_{R_{2}})=E_{\Lambda_{1}}(\boldsymbol{r}_{L_{1}}),\\
  E_{\Lambda_{1}}(\boldsymbol{r}_{R1})=-E_{\Lambda_{2}}(\boldsymbol{r}_{L_{2}}),
  \end{align}
\end{subequations}
where the MCD boundary \(L_1,L_2,R_1,R_2\) in Fig. \ref{fig2}(b) are represented by \(\boldsymbol{r}_{L_1}\),\(\boldsymbol{r}_{L_2}\), \(\boldsymbol{r}_{R_1}\), \(\boldsymbol{r}_{R_2}\) respectively,  and $\Lambda_{1}$, $\Lambda_{2}$ denotes the generated modal pair.

Therefore, from the Eq.~(\ref{eqn6}) and Eqs.~(\ref{eqn7}), we acquire the boundary conditions associated with the glide symmetry. With the natural boundary condition, which in this example is the Bloch boundary condition, corresponding to the \(\{{E}|\boldsymbol{R}_n\}\) operator.
\subsection{Finite element implementation for modal analysis}
We illustrate the basic idea of implementing NSA-FEM using the non-symmorphic symmetric structure shown in Fig. \ref{fig2}, especially the underlying principle and technical details of handling the reduced system matrix and dependent variables, as well as how are related with the counterparts of the original problems. Our target is to calculate the eigen-modal fields associated with the high symmetry $k$-points in FBZ, as marked in the band structure (see Fig. \ref{fig2}(b)) calculated from our approach compared to standard FEM in this section. The two different approach utilize nearly identical mesh density, i.e., grids per wavelength. In Fig. \ref{fig2}(c), we plot the two lowest bands along the loop of \(\Gamma-X-M-Y\) on the edge of the FBZ, i.e.,  the straight lines representing the calculation results of standard FEM and the point donating results obtained from NSA-FEM, which show perfect agreement with each other. Fig. \ref{fig2}(d) shows the non-degenerate modes \(\psi_1\) and \(\psi_2\) at k=(1/4,0). According to the Eq.~(\ref{eqn4}) and Eq.~(\ref{eqn5}), the full modal field can be obtained by applying the glide operation to the modal field associated with the MCD multiplied by the character  \(\chi(\{R|\tau+{R_{n}}\}\). In comparison, Fig. \ref{fig2}(e) shows that the degenerate modes \(\psi_{1,2}\) at \(k^*=(1/2,1/4)\) and \(k^*=(1/2,-1/4)\) transformed into each other under \(T\) symmetry.  Evidently from the comparison in Fig. \ref{fig2}(d-e) , the  eigenmode  \(\psi_1\) or \(\psi_2\) in non-degenerate case is glide symmetric by itself, while in degenerate case the eigenmodes \(\psi_1\) and  \(\psi_2\) form an inseparable  modal-pair, each of which  is transformed into its modal-pair parter and vice versa. In the following, we will give a detailed derivation of how the non-symmorphic symmetry are embedded in our NSA-FEM, especially for boundary conditions. 

For implementation details, we first execute the standard FEM procedure of eigen-value problem \cite{2002The} which will lead to matrix form of Eq.~(\ref{eqn3}),
\begin{equation}\label{eqn3} 
Ax=\lambda Bx,
\end{equation}
where \(\lambda\)  is the eigenvalue corresponding to the frequency, A and B are system matrices, and $x$ is the eigenvector with the corresponding eigenvalue $\lambda$. In combination with glide symmetry and the boundary constraints to calculate non-degenerate modes of sub-task, Eq.~(\ref{eqn3}) can be reformulated as follows,
\begin{equation}\label{eqn_add1} 
\left ( \begin{array}{cc}
    A_{1,1}&\\
     & A_{1,2}
\end{array} \right )  \left ( \begin{array}{c}
    x_{1,1}\\
   x_{1,2}
\end{array} \right )   =\lambda \left ( \begin{array}{cc}
    B_{1,1}&\\
     & B_{1,2}
\end{array} \right ) \left ( \begin{array}{c}
    x_{1,1}\\
   x_{1,2}
\end{array} \right ),
\end{equation}
where $x_{1,1}/x_{1,2}$ are the reduced eigenvectors from the original eigenvectors $x$ by glide symmetry decomposition associated with the same eigen-mode. As such, We can realize the decomposition of the original problem by calculating the decoupled sub-matrix, $ A_{1,1}x_{1,1} =\lambda B_{1,1}x_{1,1}$ or $ A_{1,2}x_{1,2} =\lambda B_{1,2}x_{1,2}$, 
and recover the original field  via the following relation between $x_{1,1}$ and $x_{1,2}$,
\begin{equation}\label{eqn_add3}
    x_{1,2}=D^{kj}(\{R|\boldsymbol{\tau+R_n}\})\{R|\boldsymbol{\tau+R_n}\}^{-1}x_{1,1}.
\end{equation}
For example, in Fig. \ref{fig2}(d), through Eq.~(\ref{eqn_add3}), we can get the full modal field recovering from the field in the red box. Moreover, from the boundary constraints in the preceding subsection, i.e., Eq.~(\ref{eqn6}), constraints on the DOFs can be derived as follows,
\begin{equation}\label{eqn8} 
{x_{1,1}}=Px_{1,1}^{\prime},
\end{equation}
where \(x_{1,1}=(x_{MCD},x_L,x_R)^T\), \(P=
\begin{pmatrix}
I & 0 & 0 \\
0 & I & \pm e^{i\pi a}I
\end{pmatrix}^T\) is the transformation matrix, \(x_{1,1}^{\prime}=(x_{MCD},x_{L})^{T}\) is the modified DOFs vector, the subscript MCD/L/R represents the DOFs associated with MCD, the left and right boundary of the MCD respectively, $I$ is the identity matrix, \(\pm e^{i\pi a}\) coefficients corresponding to the character. By using matrix operations and Eq.~(\ref{eqn8}), we can reformulate  $ A_{1,1}x_{1,1} =\lambda B_{1,1}x_{1,1}$  as, 
\begin{equation}
\label{eqn9}
    P^TA_{1,1}Px_{1,1}^{\prime}=\lambda P^TB_{1,1}Px_{1,1}^{\prime}.
\end{equation}
Furthermore, if there exists degeneracy caused by time-reversal symmetry \(T\), Eq.~(\ref{eqn3}) can be written in the following form,
\begin{widetext}
\begin{equation}\label{eqn_add4} 
\left ( \begin{array}{cccc}
    A_{1,1}&O'& &\\
    O' & A_{2,1}& &\\
    & & A_{1,2} &O'\\
    & & O' & A_{2,2}
\end{array} \right ) \begin{pmatrix}
    x_{1,1}\\
    x_{2,1}\\
    x_{1,2}\\
    x_{2,2}
\end{pmatrix} =\lambda \left ( \begin{array}{cccc}
    B_{1,1}&O'& &\\
    O' & B_{2,1}& &\\
    & & B_{1,2} &O'\\
    & & O' & B_{2,2}
\end{array} \right ) \begin{pmatrix}
    x_{1,1}\\
    x_{2,1}\\
    x_{1,2}\\
    x_{2,2}
\end{pmatrix},
\end{equation}
\end{widetext}
where $x_{1,1}/x_{1,2}$ and $x_{2,1}/x_{2,2}$ are the reduced eigenvectors from the original eigenvectors $x_1$ and $x_2$ corresponding to the Mode $\psi_1$ and $\psi_2$. Similarly, we can calculate the decoupled sub-matrix, 
\begin{equation}\label{eqn_add5} 
    \begin{pmatrix}
        A_{1,1}&O'\\ O' &A_{2,1}
    \end{pmatrix}\begin{pmatrix}
        x_{1,1}\\x_{2,1}
    \end{pmatrix}=\lambda \begin{pmatrix}
        B_{1,1}&O'\\ O' &B_{2,1}
    \end{pmatrix}\begin{pmatrix}
        x_{1,1}\\x_{2,1}
    \end{pmatrix},
\end{equation}
where the block matrices $O'$ are the extremely sparse, which denote the weak coupling terms among the diagonal block matrices. It should be noted that the eigenvector to be calculated consists of two parts of modified eigenvectors, $x_{1,1}$ and $x_{2,1}$. This is due to the fact that operations $E$ and $G_x$ both have second-order representation at $\Lambda$, thus $D^{\Lambda}$ correlates the modified eigenvectors $x_{1,1}$ and $x_{2,1}$ of the degeneracy modes. For example, in Fig. \ref{fig2}(e), $\psi_1$ and $\psi_2$ form an inseparable modal-pair. In this case, Eq.~(\ref{eqn8}) can be written in the following form, 
\begin{equation}\label{eqn_add6} 
\begin{pmatrix}
        x_{1,1}\\x_{2,1}
    \end{pmatrix}=P\begin{pmatrix}
        x_{1,1}^{\prime}\\x_{2,1}^{\prime}
    \end{pmatrix},
\end{equation}
Similarly, using matrix operations and Eq.~(\ref{eqn_add6}), Eq.~(\ref{eqn_add5}) can be written in the following form,
\begin{widetext}
\begin{equation}\label{eqn10} 
P^T\begin{pmatrix}
A_{\mathrm{MCD_1}} &  &  & & & \\
 & A_{\mathrm{L_1}} &  & & O' & \\
 &  & A_{\mathrm{R_1}} & & & \\
 & & & A_{\mathrm{MCD_2}} &  &  \\
 & O' & &  & A_{\mathrm{L_2}} &  \\
 & & &  &  & A_{\mathrm{R_2}}
\end{pmatrix}P\begin{pmatrix}
        x_{1,1}^{\prime}\\x_{2,1}^{\prime}
    \end{pmatrix}=\lambda P^T\begin{pmatrix}
B_{\mathrm{MCD_1}} &  &  & & & \\
 & B_{\mathrm{L_1}} &  & & O' & \\
 &  & B_{\mathrm{R_1}} & & & \\
 & & & B_{\mathrm{MCD_2}} &  &  \\
 & O' & &  & B_{\mathrm{L_2}} &  \\
 & & &  &  & B_{\mathrm{R_2}}
\end{pmatrix}
P\begin{pmatrix}
        x_{1,1}^{\prime}\\x_{2,1}^{\prime}
    \end{pmatrix},
\end{equation}
\end{widetext}
where \(\begin{pmatrix}
        x_{1,1}^{\prime}\\x_{2,1}^{\prime}
    \end{pmatrix}=(x_{MCD_{1}},x_{L_{1}},x_{MCD_{2}},x_{L_{2}})^{T}\),  \(\begin{pmatrix}
        x_{1,1}\\x_{2,1}
    \end{pmatrix}=(x_{MCD_{1}},x_{L_{1}},x_{R_{1}},x_{MCD_{2}},x_{L_{2}},x_{R_{2}})^{T}\),
\(P=
\begin{pmatrix}
I & 0 & 0 & 0 \\
0 & I & 0 & 0 \\
0 & 0 & 0 & -I \\
0 & 0 & I & 0 \\
0 & 0 & 0 & I \\
0 & I & 0 & 0
\end{pmatrix}\) is the transformation matrix, 
the subscript MCD/\(L_i\)/\(R_i\) represents the coefficients within the calculation domain/on the MCD boundary \(L_i\)/on the MCD boundary \(R_i\), \(I\) is the identity matrix,\(-1\) coefficients corresponding to the character.  Eq. (\ref{eqn10}) is the essentially the system matrix  that needs to be solved, and $(x_{1,1}^{\prime}, x_{2,1}^{\prime})$ is the dependent variable that can be used to recover the full solution in junction with the transformation matrix $P$.

\section{Results and discussion}
We proceed to validate the effectiveness and validity of our method through three concrete examples involving a subperiodic group(rod groups), a plane group and a space group, which are compared with the standard FEM. Our method uses a mesh density consistent with the standard FEM and employs the same numerical solver $eigs$ in MATLAB. The computer configuration includes a 12th Gen Intel Core i5-12400 processor and 16 GB of RAM. All numerical results were generated using homemade MATLAB code. The MATLAB implementation of our NSA-FEM procedure is open-source and available on Github \cite{GitHub}.

\subsection{AB-layer-stacked photonic crystal with layer group}
\begin{figure}
\includegraphics{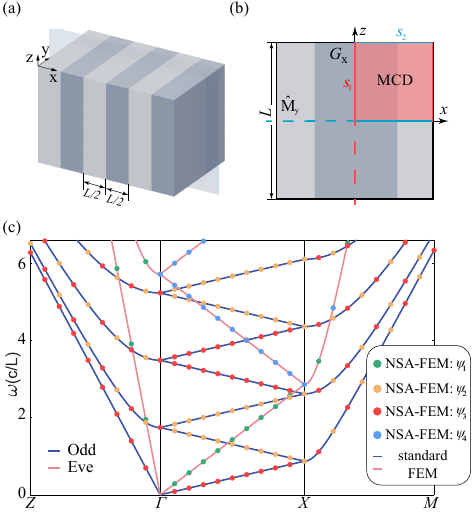}
\caption{\label{fig:epsart} (a) The structure of the photonic crystal's unit cell in 3D, consists of stacked biaxial dielectrics in layer A (light gray) and layer B (dark gray). (b) The  MCD and symmetry axis in computational plane xz. (c) Band structure along high symmetry lines in the ky = 0 plane for the PHC. }
\label{fig3}
\end{figure}

In previous study, AB-layer-stacked photonic crystal (Phc)  manifests a hidden symmetry of Maxwell’s equations, reveals a novel mechanism to realize protected degeneracies unique to photonic bands. The structure of the Phc considered here shows in Fig. \ref{fig3}(a), consists of two types of dielectric layer (A and B), which have equal thicknesses L/2 and stacked periodically along the x direction\cite{2020Hidden}. As such, the PhC’s relative permittivity tensor in xyz coordinates is given by
\begin{equation}\label{eqn11}
\varepsilon_r=
\begin{pmatrix}
\varepsilon_{xx} & 0 & \varepsilon_{xz} \\
0 & \varepsilon_{yy} & 0 \\
\varepsilon_{zx} & 0 & \varepsilon_{zz}
\end{pmatrix}
\end{equation}
where \(\varepsilon_{xz}=\varepsilon_{zx}=\pm g=\pm(\varepsilon_{1}-\varepsilon_{3})\mathrm{sin}\theta\mathrm{cos}\theta\) flips its sign
between layers A and B, while the diagonal elements \(\varepsilon_{xx}=(\cos^2\theta\varepsilon_1+\varepsilon_3\mathrm{sin}^2\theta)\), \(\varepsilon_{yy}=\varepsilon_{2}\), \(\varepsilon_{zz}=(\sin^2\theta\varepsilon_1+\varepsilon_3\mathrm{cos}^2\theta)\) are all constant. Here we set \(\theta=\pi/5\), \(\varepsilon_{1}=2\), \(\varepsilon_{1}=13\), \(\varepsilon_{1}=1\).
In this model, the rod group index is pmcm(number 22), which semidirectly products with the two-dimensional continuous translational group \(\mathbb{R}^2\), i.e., \(\mathbb{R}^2\rtimes\mathrm{Rod}(22)\). Rod groups are subperiodic groups of space groups, which can be considered as discrete symmetry groups of three-dimensional objects, translationally periodic along a line. Therefore, in numerical analysis, we can use the standard FEM to simulate a two-dimensional unit cell that is periodically continuous in out-of-plane direction, as shown in Fig. \ref{fig3}(b). The red area indicates the MCD of NSA-FEM, which is one-quarter of its original size reduced by the combination of \(M_y\) and \(G_x\) symmetries. In $k$-space, here we focus on several space group symmetries relevant to band crossings in the plane \(k_y=0\). The Irreps of high symmetry momentum for the rod group pmcm with \(k_y=0\) plane are shown in Table \ref{tab:table2}.
\begin{table}
\caption{\label{tab:table2}Irreps of the high symmetry \(\boldsymbol{k}\)-vector for rod group pmcm}
\begin{ruledtabular}
\renewcommand\arraystretch{1.4}
\begin{tabular}{c|ccccc}
 Location&Seitz Symbol&\multicolumn{4}{c}{Representation}\\ \hline
\multirow{4}{*}{$\Lambda$:(0,0,$a$)\footnote{$0<a <\frac{1}{2}$}}&\(\{1|t_1,t_2,t_3\}\)&\multicolumn{4}{c}{\(e^{i2\pi t_3a}\)} \\ 
&\(\{2_{001}|\frac{1}{2},0,0\}\)& 1 & 1& -1& -1 \\
&\(\{m_{010}|0,0,0\}\)& 1 & -1& -1& 1 \\
&\(\{m_{100}|\frac{1}{2},0,0\}\)&1&-1&1&-1\\ \Xcline{1-6}{0.05pt}
\multirow{4}{*}{$\Sigma$:($a$,0,0)}&\(\{1|t_1,t_2,t_3\}\)&\multicolumn{4}{c}{\(e^{i2\pi t_1a}\)}\\ 
&\(\{2_{100}|\frac{1}{2},0,0\}\)& \(e^{i\pi a}\) & \(e^{i\pi a}\)& \(e^{i\pi (1+a)}\)& \(e^{i\pi (1+a)}\) \\
&\(\{m_{010}|0,0,0\}\)&1&-1&-1&1\\
&\(\{m_{001}|\frac{1}{2},0,0\}\)& \(e^{i\pi a}\) & \(e^{i\pi (1+a)}\)& \(e^{i\pi a}\)& \(e^{i\pi (1+a)}\) \\ \Xcline{1-6}{0.05pt}
\multirow{4}{*}{$G$:($\frac{1}{2}$,0,$a$)}&\(\{1|t_1,t_2,t_3\}\)&\multicolumn{4}{c}{\(\mathrm{e^{i\pi(t_1+2t_3a)}}\)} \\
&\(\{2_{001}|\frac{1}{2},0,0\}\)& 1 & 1& -1& -1 \\
&\(\{m_{010}|0,0,0\}\)& 1 & -1& -1& 1 \\
&\(\{m_{100}|\frac{1}{2},0,0\}\)&1&-1&1&-1\\
\end{tabular}
\end{ruledtabular}
\end{table}

At each of the three locations, \(\Lambda\), \(\Sigma\), and \(G\), there are four distinct classes, i.e., see details in the column of the Seitz symbol. As for a specific representation (column number fixed), the character indifferent row is to describe the conserved quantity of symmetric operation for different class, and thus reflects the symmetric properties of that Irreps under the symmetric operation belonging to different class. Specifically, the class consisting of the operation \(\{1|t_1,t_2,t_3\}\) corresponds to the normal Bloch boundary conditions. Each of the other three classes can be derived as the combination of the two remaining classes. For instance, the class consisting of the operation \(\{m_{100}|\frac{1}{2},0,0\}\) can be expressed as the combination of the two remaining classes consisting of the operations \(\{2_{001}|\frac{1}{2},0,0\}\{m_{010}|0,0,0\}\), the corresponding matrix transformation representation is:
\begin{equation}\label{eqn12}
    \left.\left(
\begin{array}
{rrrrr}-1 & 0 & 0 & \frac{1}{2} \\
0 & 1 & 0 & 0 \\
0 & 0 & 1 & 0
\end{array}\right.\right)=
\begin{pmatrix}
-1 & 0 & 0 & \frac{1}{2} \\
0 & -1 & 0 & 0 \\
0 & 0 & 1 & 0
\end{pmatrix}
\begin{pmatrix}
 & 1 & 0 & 0 & & 0 \\
 & 0 & -1 & 0 & & 0 \\
 & 0 & 0 & 1 & & 0
\end{pmatrix}
\end{equation}
To effectively decompose the problem and apply symmetry constraint boundary conditions, only two independent symmetric operations are required: mirror symmetry \(M_y\) and the glide symmetry \(G_x=\{M_x|(a/2)\hat{x}\}\). The symmetries \(M_y\) and \(G_x\) both allow the computation domain to be reduced by half. Consequently, the combination of the \(M_y\) and \(G_x\) symmetries enable the structure to be reduced to one-quarter of its original size. And since each location possesses four distinct Irreps, we can further divide the original problem into 4 sub-tasks.

The band structure calculation is performed along the boundaries of the irreducible FBZ. And the calculation results are illustrated in Fig. \ref{fig3}(c), where the straight lines represent the calculation results of standard FEM and the point results obtained from NSA-FEM. The blue and magenta lines represent the bands with odd and even \(\hat{M}_{y}\)-parities, respectively. From the perspective of mode analysis,  the generation of even and odd modes is attributed to \(M_y\) symmetry. Therefore, modes can be divided into two types based on \(\chi(\{M_y|0\})=\pm1\), corresponding to \(\psi_{1,4}\) and \(\psi_{2,3}\), which represent even and odd modes, respectively.
 From the computational perspective, according to Equation: \(\boldsymbol{E}(\boldsymbol{r})=\left[\boldsymbol{E}_t(x,y)+\boldsymbol{\hat{z}}\boldsymbol{E}_z(x,y)\right]e^{-\gamma z}\), we can calculate even and odd modes separately, effectively distinguishing them while ensuring that the higher-order degenerate nexus points remain intact, which significantly enhances the efficiency of band structure calculations. The computation time for the standard FEM is 1,067 seconds. In contrast, the NSA-FEM significantly reduces this time to just 320 seconds, with the four sub-tasks taking 51 seconds, 94 seconds, 115 seconds, and 60 seconds, respectively. The DOFs for the standard FEM problem are 9,841, with an average memory usage of 2.38 GB, while the NSA-FEM problem has 2,521 degrees of freedom and an average memory usage of 2.08 GB. Compared to the standard finite element method, the DOFs are reduced to one-fourth, the memory usage is reduced by 12.6\(\%\), and the computation time is reduced to 30.0\(\%\), which demonstrates the effectiveness of our method.

\subsection{Twisted Quadrupole Topological Photonic Crystals with plane group}
\begin{figure}
\includegraphics{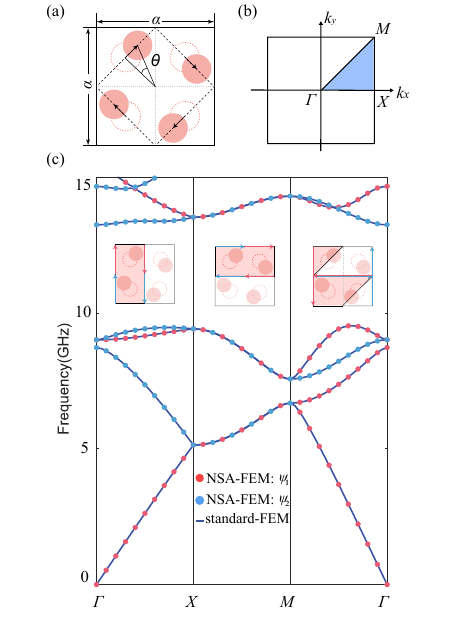}
\caption{\label{fig:epsart} (a) Structure of the twisted quadrupole topological photonic crystals, twisting is along the dashed lines with \(\theta=22^{\circ}\). (b) The FBZ and high symmetry momentum (\(k\)-vectors) in \(k\)-space where the blue region represents irreducible FBZ.  (c) Photonic band structure along \(\Gamma-X-M-\Gamma\).}
\label{fig4}
\end{figure}
In the second example, recent development in topological photonics shows that lateral hetero structures with different twisting angles can yield rich topological phenomena, and thus opens a new route toward topological photonics\cite{https://doi.org/10.1002/lpor.202000010}. The 2D square-lattice PhC has four identical cylinders made of $Al_2O_3$ (radius 
\(r=0.25cm\), \(\varepsilon_{r}=6.2,\mu_{r}=1\) ) in each unit cell. The lattice constant is
\(a=2cm\). By twisting the four cylinders along the dashed lines, which are initially located at the positions of \((\pm\frac{a}{4},\pm\frac{a}{4})\), as illustrated in Fig. \ref{fig4}(a), the symmetry is reduced to the nonsymmorphic plane group P4g(number 12), which contains the two glide symmetries. In $k$-space, the FBZ and the high symmetry $k$-vectors are shown in Fig. \ref{fig4}(b), where the blue region represents irreducible FBZ and the band analysis is along the boundary of irreducible FBZ, i.e., \(\Gamma-X-M-\Gamma\). The Irreps of high symmetry $k$-vectors are shown in Table \ref{tab:table3}.

\begin{table}
\caption{\label{tab:table3}Irreps of the high symmetry \(\boldsymbol{k}\)-vector for plane group P4bm}
\begin{ruledtabular}
\renewcommand\arraystretch{1.4}
\begin{tabular}{c|ccc}
 Location&Seitz Symbol&\multicolumn{2}{c}{Representation}\\ \hline
\multirow{2}{*}{$\Delta$:($a$,0)\footnote{$0<a <\frac{1}{2}$}}&\(\{1|t_1,t_2\}\)&\(e^{i2\pi t_1a}\)&\(e^{i2\pi t_1a}\) \\ 
&\(\{\mathrm{m}_{01}|\frac{1}{2},\frac{1}{2}\}\)& \(e^{i\pi a}\) & \(e^{i\pi (1+a)}\) \\ \Xcline{1-4}{0.05pt}
\multirow{2}{*}{$\Lambda$:($\frac{1}{2}$,$a$)}&\(\{1|t_1,t_2\}\)&\(e^{i\pi (t_1a+2t_2a)}\)&\(e^{i\pi (t_1a+2t_2a)}\) \\ 
&\(\{\mathrm{m}_{10}|\frac{1}{2},\frac{1}{2}\}\)& \(e^{i\pi a}\) & \(e^{i\pi (1+a)}\) \\ \Xcline{1-4}{0.05pt}
\multirow{2}{*}{$\Sigma$:($a$,$a$)}&\(\{1|t_1,t_2\}\)& \(\mathrm{e^{i2\pi(t_1a+t_2a)}}\)& \(\mathrm{e^{i2\pi(t_1a+t_2a)}}\) \\
&\(\{\mathrm{m}_{1\overline{1}}|\frac{1}{2},\frac{1}{2}\}\)& \(\mathrm{e}^{\mathrm{i}2\pi a}\)& \(\mathrm{e}^{\mathrm{i}2\pi (1+a)}\) \\
\end{tabular}
\end{ruledtabular}
\end{table}
In the numerical analysis of the band structure shown in Fig. \ref{fig4}, we only calculate the eigen modes with dominating the out-of-plane field component, since out-of-plane field components of eigen modes at low frequencies dominate. It should be noted that, compared to the first example, the Seitz symbols corresponding to the glide operation are different at locations \(\Delta\), \(\Lambda\), and \(\Sigma\), which means that we shall truncate the full unit cell differently into different MCDs, as shown in three insets in Fig. \ref{fig4}(c). The insets are column-wise corresponding to the band $\Gamma-X$, $X-M$, $M-\Gamma$, respectively.  The rationale is the following: the symmetry operations associated with the three different Seitz symbols at band $\Gamma-X$, $X-M$, $M-\Gamma$ are different. 
\begin{enumerate}
    \item At \(\Delta\), the Seitz symbol \(\{\mathrm{m}_{010}|\frac{1}{2},\frac{1}{2},0\}\) reduces the MCD to the left half of the unit cell. In the upper left quarter, the red boundaries constraints correspond inversely after the glide operation, with a similar effect in the lower left quarter.
    \item At \(\Lambda\), the Seitz symbol \(\{\mathrm{m}_{100}|\frac{1}{2},\frac{1}{2},0\}\)  reduces the MCD to the upper half of the unit cell, with boundary constraints correspondence similar to \(\Delta\).
    \item 
    At \(\Sigma\), the MCD within the unit cell is not regular; instead, it is reduced along the symmetry axes \(y=x+\frac{1}{2}\) and \(y=x+-\frac{1}{2}\) to half. By effectively translating the eighth triangle of the upper left along \(\Delta y =-a\) to the next cell, it is evident that the computational region is reduced to a triangle.
\end{enumerate}
In conclusion, the magenta region highlights the MCD, and the red and blue boundaries indicate the boundary constraint relations and the matching directions, as denoted by the two red-blue arrow pairs. Based on the symmetry property of glide symmetry, we can divide the original problem into two sub-tasks.

As shown in Fig. \ref{fig4}(c), the straight lines represent the results calculated by standard FEM, while the points denote those calculated using NSA-FEM. The results obtained through our method are consistent with those from the standard FEM. In addition to improved computational efficiency, our method facilitates the effective distinction of modes across different sub-tasks, simplifying the differentiation between adjacent bands. In this model, the computation time for the standard FEM is 319 seconds, whereas the total computation time for the NSA-FEM is 242 seconds. The DOFs for the standard FEM problem are 156,217, compared to 76,415 DOFs for the NSA-FEM. It is noteworthy that, although Model B features a larger number of DOFs compared to Model A, the reduction in computation time is limited. This is attributed to several factors: the number of eigenvalues to be solved is smaller, the eigenvalue equation differs in its solution difficulty, and the model only exhibits glide symmetry without combined symmetries. In summary, compared to the standard FEM, the NSA-FEM reduces the DOFs by half and decreases the computation time to 75.9\(\%\).

\subsection{Photonic nodal line with space group}
\begin{figure}
\includegraphics{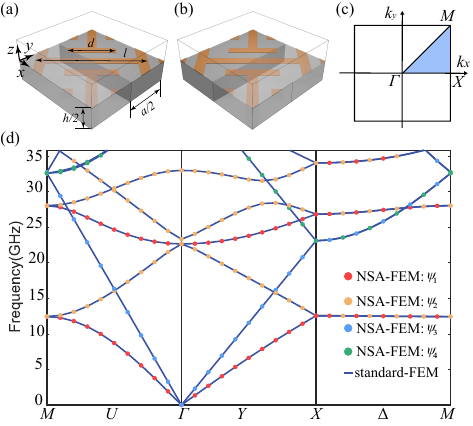}
\caption{\label{fig:epsart} (a) MCD at crystal momentum U and Y. (b) MCD at crystal momentum \(\Delta\). (c) The FBZ and high symmetry momentum (\(k\)-vectors) in \(k\)-space where the blue region represents irreducible FBZ. (d) Band structure of the metacrystal along \(M-\Gamma-X-M\). Nodal line (NL) is formed between the second and third band at 16 GHz at crystal momentum U and Y. }
\label{fig5}
\end{figure}
Nodal line semimetals (NLs) are three-dimensional (3D) crystals that support band crossings in the form of one-dimensional rings in the Brillouin zone. In the presence of spin–orbit coupling or lowered crystal symmetry, NLS may transform into Dirac semimetals, Weyl semimetals, or 3D topological insulators\cite{2018Experimental}. We perform a band structure analysis on the metacrystal. By introducing glide symmetry into the cut-wire meta material design, realizing a negative dispersion for the LP mode, which plays a key role in the formation of NL degeneracy in this model. The unit cell metacrystal, with a dimension of \(4.5\times4.5\times2\mathrm{mm}^{3}\) , consists of two mutually orthogonal I-shaped metallic cut-wire resonators lying in the x–y plane. In Fig. \ref{fig5}(a,b), the solid line region shows that the unit cell size of $a = 4.5 mm$, $h = 2 mm$. The size of the I-shaped copper cut-wire resonators is $l = 1.1a$, $d = 0.5a$. The widths of all wires are $0.1a$, and the thicknesses are  \(35 \mu m\). The substrate material is Teflon, whose permittivity is 2.1, with a loss tangent around 0.00028. The dark gray region indicates the MCD at crystal momentum $U$, $Y$, and $\Delta$. The metacrystal space group index is P4/mbm (number 127), which exhibits two glide symmetries perpendicular to the main axis. In $k$-space, the FBZ and the high symmetry $k$-vectors are shown in Fig. \ref{fig5}(c), where the blue region represents irreducible FBZ and the band analysis is along the boundary of irreducible FBZ, i.e., \(M-\Gamma-X-M\). The Irreps of high symmetry momentum are shown in Table \ref{tab:table4}. 

Similarly to model A, only two independent cases can be used: Based on the symmetry properties of glide symmetry $G$ and mirro symmetry $M$ , we can divide the original problem into four sub-tasks \(\psi_1,\psi_2,\psi_3\), and \(\psi_4\). Fig. \ref{fig5}(d) illustrates the band structure of the metacrystal along \(M-\Gamma-X-M\). The straight lines represent results calculated by standard FEM, while the points denote those calculated using NSA-FEM. We compared the MCD, the number of degrees of freedom (DOFs), and the number of tasks between NSA-FEM and standard finite element modeling. Although NSA-FEM requires calculating four sub-tasks separately, the total calculation time is significantly reduced to 4,851 seconds, compared to 23,366 seconds for standard FEM. The DOFs for the standard FEM problem are 1,036,485, whereas the NSA-FEM involves just 248,369 DOFs, the DOFs are reduced to one-fourth improving calculation speed by 4.8 times. Thus, with more sub-tasks, a smaller MCD, and fewer DOFs, NSA-FEM achieves superior computational efficiency.
\begin{table}
\caption{\label{tab:table4}Irreps of the high symmetry \(\boldsymbol{k}\)-vector for space group P4/mbm}
\begin{ruledtabular}
\renewcommand\arraystretch{1.4}
\begin{tabular}{c|ccccc}
 Location&Seitz Symbol&\multicolumn{4}{c}{Representation}\\ \hline
\multirow{4}{*}{$U$:($a$,$a$,0)\footnote{$0<a <\frac{1}{2}$}}&\(\{1|t_1,t_2,t_3\}\)&\multicolumn{4}{c}{\(e^{i2\pi (t_1a+t_2a)}\)} \\ 
&\(\{2_{001}|\frac{1}{2},\frac{1}{2},0\}\)& \(e^{i2\pi a}\) & \(e^{i2\pi (1+a)}\)& \(e^{i2\pi (1+a)}\)& \(e^{i2\pi a}\) \\
&\(\{m_{001}|0,0,0\}\)& 1 & 1& -1& -1 \\
&\(\{\mathrm{m}_{1\bar{1}0}|\frac{1}{2},\frac{1}{2},0\}\)&\(e^{i2\pi a}\)&\(e^{i2\pi (1+a)}\)&\(e^{i2\pi a}\)&\(e^{i2\pi (1+a)}\)\\ \Xcline{1-6}{0.05pt}
\multirow{4}{*}{$Y$:($a$,0,0)}&\(\{1|t_1,t_2,t_3\}\)&\multicolumn{4}{c}{\(e^{i2\pi t_1a}\)} \\
&\(\{2_{100}|\frac{1}{2},\frac{1}{2},0\}\)& \(e^{i\pi a}\)& \(e^{i\pi (1+a)}\)& \(e^{i\pi (1+a)}\) & \(e^{i\pi a}\) \\
&\(\{m_{001}|0,0,0\}\)&1&1&-1&-1\\
&\(\{m_{001}|\frac{1}{2},0,0\}\) & \(e^{i\pi a}\) & \(e^{i\pi (1+a)}\) & \(e^{i\pi a}\) & \(e^{i\pi (1+a)}\) \\ \Xcline{1-6}{0.05pt}
\multirow{4}{*}{$G$:($\frac{1}{2}$,$a$,0)}&\(\{1|t_1,t_2,t_3\}\)&\multicolumn{4}{c}{\(\mathrm{e^{i\pi(t_1+2t_2a)}}\)} \\
&\(\{2_{010}|\frac{1}{2},\frac{1}{2},0\}\)& \(e^{i\pi a}\)& \(e^{i\pi (1+a)}\)& \(e^{i\pi (1+a)}\) & \(e^{i\pi a}\) \\
&\(\{m_{001}|0,0,0\}\)&1&1&-1&-1\\
&\(\{m_{100}|\frac{1}{2},\frac{1}{2},0\}\)& \(e^{i\pi a}\) & \(e^{i\pi (1+a)}\) & \(e^{i\pi a}\) & \(e^{i\pi (1+a)}\) \\
\end{tabular}
\end{ruledtabular}
\end{table}

Along the in-plane directions, the lowest three bands are formed by two transverse modes, the transverse electric (TE) and the transverse magnetic (TM) modes, and a LP mode. The orthogonality between the LP and TE modes is guaranteed by the mirror symmetry of the system. Therefore, same to the model A, the modes can be divided into two types based on \(\chi(\{M|0\})=\pm1\), corresponding to \(\psi_{1,2}\) and \(\psi_{3,4}\), where TM mode corresponding to \(\psi_1\), LP mode corresponding to \(\psi_2\), TE mode corresponding to \(\psi_3\). The degeneracy of LP and TM modes is caused by the glide symmetry of the system, which can be divided according to \(\chi(\{M|\boldsymbol{\tau}\})=\pm e^{i\pi a}\).

\section{CONCLUSION}
\begin{table}
\caption{\label{tab:table5}Calculation summary of FEM and NSA-FEM, MCD, number of DOFs, tasks number and computation time}
\renewcommand\arraystretch{1.4}
\begin{ruledtabular}
\begin{tabular}{c|c|cccc}
Methods&\makecell{Structure\\Type}& MCD & DOFs & Tasks &\makecell{Computation\\Time}\\ \hline
\multirow{3}{*}{FEM}&\makecell{AB-layer\\PHC}&D&9,841&1& 1067s\\
&TQT PHC&D&156,217&1& 319s\\
&NL&D&1,036,485&1& 23,366s\\ \Xcline{1-6}{0.05pt}
\multirow{3}{*}{\makecell{NSA-\\FEM}}&\makecell{AB-layer\\PHC}& \(D/4\) &2,521& 4 & 320s\\
&TQT PHC&D/2&76,415&2& 242s\\
&NL&D/4&248,369&4& 4,851s\\
\end{tabular}
\end{ruledtabular}
\end{table}
In summary, we systematically applied non-symmorphic symmetries to the numerical analysis of band structures of photonic structure and proposed the NSA-FEM. We implemented NSA-FEM across groups of different dimensions, including subperiodic groups, plane groups, and space groups. Based on the Irreps of groups, we decompose the original problem into several sub-tasks, each corresponding to different symmetry type of modes, while ensuring that their topological properties remain intact. Meanwhile, the decomposition of sub-tasks also introduces the potential for parallelization in the algorithm. Table \ref{tab:table5} shows the summary of the calculation of FEM and NSA-FEM, compared to the standard FEM, our method not only significantly improves the efficiency of the analysis of the band structure, but also facilitates the discovery of hidden symmetries, high-order degenerate points, nodal line degeneracies, and other novel topological phenomena based on their symmetry types. Additionally, by dividing modes into different subtasks and applying character constraints corresponding to the Irreps to calculate the corresponding modes, we provide a perspective for understanding and computing these modes. Our work fills the gap in using nonsymmorphic symmetries to reduce computational domains, enabling researchers in optics, condensed matter, and numerical analysis to systematically apply group theory in FEM to support their research efforts. Considering the fundamental role of character tables in group theory, our method is expected to find broader applications in other groups, such as double groups and magnetic groups. Additionally, due to the universality of group theory, this symmetry-based approach is versatile and can be easily extended from FEM to other numerical techniques, such as the FDM and the transfer matrix method.

\begin{acknowledgments}
We thank Dr. Ruo-Yang Zhang for fruitful discussions. This work was supported by the National Natural Science Foundation of China (Grants No. 12274161), the Optics Valley Laboratory, Wuhan, Hubei 430074, China and the Natural Science Foundation of Hubei Province of China under Grant 2024AFA016.
\end{acknowledgments}

\appendix

\section{Normal-subgroup induction method for the calculation of space groups Irreps}
\label{ap1}

The calculation of Irreps of space groups follows the algorithm based on a normal-subgroup induction method. The process can be divided into six steps:
 \begin{enumerate}
    \item Identify the second kind subgroup of the space group, known as the little group \(G_k\).
    \item Determine the kernel group \(T_k\) of \(G_k\), which is the invariant subgroup with the character equal to 1.
    \item Calculate the characters of the Irreps of the quotient group \(G_k/T_k\).
    \item Generate the Irreps of \(G_k\) from the Irreps of \(G_k/T_k\).
    \item Identify the allowed representations of \(G_k\).
    \item Induce the Irreps of the space group from the allowed representations.
\end{enumerate}
Take the example of solving the Irreps of the plane group pg at the Location X(\(\boldsymbol{k}=\frac{\pi}{a}\boldsymbol{i}\)):

Step 1: Identify the second kind subgroup:

The second kind subgroup is given by
\begin{equation}
\label{apeqn1}
    G_k=\{E|0\}T+\{\sigma_x|\boldsymbol{\tau}\}T,
\end{equation}
where \(\sigma_x\) denotes a mirror reflection plane and \(\tau\) is the translation associated with this reflection.
     
Step 2: Determine the kernel group \(T_k\):

The kernel group \(T_k\)  is an invariant subgroup with a character equal to 1, which implies that the elements \(\{E|\boldsymbol{R}_n\}\) of \(T_k\) satisfy condition \(e^{i\boldsymbol{kR_n}}=1\). Given \(\boldsymbol{k}=\frac{\pi}{a}\boldsymbol{i}\), this condition requires \(\boldsymbol{R}_n=2ma_1\boldsymbol{i}\), where \(m\) is an integer and \(a_1\) is the lattice constant along the x direction. Thus, the kernel group is expressed as:
\begin{equation}
    T=T_k\{E|0\}+T_k\{E|\boldsymbol{a_1}\}.
\end{equation}
    
Step 3: Calculate the characters of the Irreps of the quotient group:

The quotient group is
\begin{equation}
    G_k/T_k=\{E|0\}+\{E|\boldsymbol{a}_1\}+\{\sigma_x|\boldsymbol{\tau}\}+\{\sigma_x|\boldsymbol{\tau+a}_1\}.
\end{equation}
This indicates that the quotient group consists of four elements, specifically \(A_1:\{E|0\}\), \(A_2:\{E|\boldsymbol{a}_1\}\), \(A_3:\{\sigma_x|\boldsymbol{\tau}\}\), \(A_4:\{\sigma_x|\boldsymbol{\tau+a}_1\}\). Their multiplication table is shown in \ref{tab:table6}. 

\begin{table}
\caption{\label{tab:table6}Multiplication table for elements of the quotient group} 
\begin{ruledtabular}
\renewcommand\arraystretch{1.4}
\begin{tabular}{ccccc}
&\(A_1\) &\(A_2\) &\(A_3\)&\(A_4\)  \\ \hline
\(A_1\)&\(A_1\) &\(A_2\) &\(A_3\)&\(A_4\)  \\
\(A_2\)&\(A_2\) &\(A_1\) &\(A_4\)&\(A_3\)  \\
\(A_3\)&\(A_3\) &\(A_4\) &\(A_2\)&\(A_1\)  \\
\(A_4\)&\(A_4\) &\(A_3\) &\(A_1\)&\(A_2\)  \\
\end{tabular}
\end{ruledtabular}
\end{table}
This group is isomorphic to the cyclic group of order four, \(G^1_4\), and its character table is presented in Table \ref{tab:table7}. All the important properties associated with the symmetry group can be extracted from the character table:
\begin{enumerate}
\item The character table lists all the Irreps and the characters row-by-row, i.e., which essentially corresponds to different symmetric type of modal in our paper; 
\item Each column represents a class of the group; 
\item As for a specific irreducible representation (row number fixed), the character indifferent column is to describe the conserved quantity of matrix representation for different class, and thus reflects the symmetric properties of that irreducible representation under the symmetric operation belonging to different class.
\end{enumerate}

\begin{table}
\caption{\label{tab:table7}The character table} 
\begin{ruledtabular}
\renewcommand\arraystretch{1.4}
\begin{tabular}{ccccc}
Label&\(\{E|0\}\)&\(\{\sigma_x|\boldsymbol{\tau}\}\)&\(\{E|\boldsymbol{a}_1\}\) &\(\{\sigma_x|\boldsymbol{\tau+a}_1\}\)  \\ \hline
\(X_1\)&1 &1 &1&1  \\
\(X_2\)&1 &\(i\) &-1&\(-i\)  \\
\(X_3\)&1 &-1 &1&-1  \\
\(X_4\)&1 &\(-i\) &-1&\(i\)  \\
\end{tabular}
\end{ruledtabular}
\end{table}

Step 4 \& 5: Generate and identify the allowed representations of \(G_k\)

From the Bloch boundary conditions, we obtain
\begin{equation}
    \chi\{E|\boldsymbol{a}_1\}=e^{-i\boldsymbol{ka}_1}=-1,
\end{equation}
implying that \(\chi\{\sigma_x|\boldsymbol{\tau}\}=\pm i\). Consequently, only \(X_2\) and \(X_4\) are the allowed representations of \(G_k\).

Step 6: Induce Irreps of the space group.

For different \(\boldsymbol{k}\)-vectors, applying the above operations allows us to derive the Irreps of the space group induced by the allowed representations.

\section{The arms of the star $\boldsymbol{k}*$ induction method for the calculation of full-group Irreps}
\label{ap2}

The indication of the arms of the star $\boldsymbol{k}*$ precedes the table of the matrices of the full-group Irreps $D^{\boldsymbol{k}*,j}$ of $^{d}G$ induced from the allowed irreps $D^{\boldsymbol{k},j}$ of the little group $^{d}G^k$. As shown in Table \ref{tab:table8} and Table \ref{tab:table9}, the structure and organization of the data of the full-group Irreps table are very similar to that of the little-group irreps: The coset representatives of the decomposition $^{d}G:T$ of the space group with respect to the translation subgroup specify the rows of the table, while the columns correspond to the full space-group irreps. Symmetry operations are described by their matrix–column pairs and Seitz symbols. The full Irreps are of dimension $r\times s$, where $s$ is the number of arms of the star $k*$ and $r$ is the dimension of the corresponding allowed Irrep of $^{d}G^{k}$. The matrices of the full Irreps have a block structure with $s\times s$ blocks, each of dimension $r\times r$.

\begin{table}
\caption{\label{tab:table8}Matrices of the representations of the little group at location $\Lambda$} 
\begin{ruledtabular}
\renewcommand\arraystretch{1.4}
\begin{tabular}{ccc}
Matrix presentation&Seitz Symbol&$A_1$  \\ \hline
$\begin{pmatrix} & 1 & & 0 &    & \mathrm{t}_1 \\ & 0 & & 1 &   & \mathrm{t}_2 \\ \end{pmatrix}$&$\{1|t_1,t_2\}$ &$e^{i\pi(t_1+2t_2a)}$\\
\end{tabular}
\end{ruledtabular}
\end{table}

\begin{table}
\caption{\label{tab:table9}Matrices of the representations of the full-group at location $\Lambda$} 
\begin{ruledtabular}
\renewcommand\arraystretch{1.4}
\begin{tabular}{ccc}
Matrix presentation&Seitz Symbol&$^{*}A_1(-1)$  \\ \hline
$\begin{pmatrix} & 1 & & 0 &    & \mathrm{t}_1 \\ & 0 & & 1 &   & \mathrm{t}_2 \\ \end{pmatrix}$&$\{1|t_1,t_2\}$ &{\(\left.\left(
\begin{array}
{cc}e^{i\pi(t_1+2t_2a)} & 0 \\
0 & e^{i\pi(t_1+2t_2a)}
\end{array}\right.\right)\)}\\
$\begin{pmatrix} & 1 & 0 & & \frac{1}{2} \\ & 0 & -1 && 0  \end{pmatrix}$&\(\begin{array} {c}\{\mathfrak{m}_{01}|\frac{1}{2},0\} \end{array}\) &{\(\left.\left(
\begin{array}
{cc}0 & 1 \\
-1 & 0
\end{array}\right.\right)\)}\\
\end{tabular}
\end{ruledtabular}
\end{table}

For applications involving time-reversal symmetry, it is necessary to give a brief overview of the Irreps classification of complex conjugation or, as it is commonly referred to, with respect to their 'reality'. The number in brackets after the Irreps label in Table \ref{tab:table9} specifies the reality of the Irreps as per equation:
\begin{equation}
    \left.\frac{1}{|\mathcal{G}|}\sum_{j=1}^{|\mathcal{G}|}\chi\left(g_j^2\right)=\left\{
\begin{array}
{cc}1 & \text{iff D is of the first kind,} \\
-1 & \text{iff D is of the second kind,} \\
0 & \text{iff D is of the third kind,}
\end{array}\right.\right.
\end{equation}

(1) indicates an Irrep of the first kind, i.e. real; (-1) an Irrep of the second kind, or pseudo-real; and (0) an Irrep of the third kind, or complex. The Irreps of the first and second kind are also known as 'self-conjugate', while the Irreps of the third kin form pairs of conjugated Irreps $(D,D^*)$ which, in general, may be induced from allowed Irreps of wave vectors belonging to different stars. Once the reality of a space group Irrep has been determined, the time-reversal (TR)-invariant Irreps can be constructed according to the following criteria:
 \begin{enumerate}
    \item If Irrep $D$ is (a) single valued and real or (b) double valued and pseudo-real, it is TR invariant.
    \item If Irrep $D$ is (a) single valued and pseudo-real or (b) double valued and real, the TR-invariant Irrep is the direct sum of $D$ with itself. The dimension of the TR-invariant Irrep doubles the dimension of $D$. The label of the TR-invariant Irrep consists of two copies of the label of $D$.
    \item If $D_1$ and $D_2$ form a pair of mutually conjugated Irreps, the direct sum of both Irreps is TR invariant. The label of the TR-invariant representation is the union of the labels of the two Irreps.
\end{enumerate}

\bibliography{apssamp}

\end{document}